# THE DETECTION OF DENIS GALAXIES

V. BANCHET[1], G.A. MAMON[1,2] & M. CONTENSOU[2]
[1] Institut d'Astrophysique de Paris, F–75014 Paris, FRANCE
[2] DAEC, Observatoire de Paris, F–92195 Meudon, FRANCE

**Abstract** – Using simulated images, we test various methods, for the homogeneous detection of objects in DENIS images. The optimal detection limits depend on the object type and the sky background, but little on the PSF or the smoothing filter used to increase the signal-to-noise of the images.

## 1. INTRODUCTION

The task of building a roughly homogeneous catalog (for cosmological applications) of galaxies from the DENIS database of images (see Fouqué, in these proceedings) is hampered by the intrinsic difficulty to detect galaxies with DENIS, as the $K'$-band surface brightness of galactic disks viewed face-on is much below the background. Moreover, we expect only a few at most galaxies per $12' \times 12'$ image, with at least 100 times as many stars (over $10^4$ times more at low galactic latitude). Our goal is to find a rapid method for a homogeneous detection of an optimal fraction (the *selection function*) of galaxies to a given magnitude limit and with a given confidence level. We use simulated images for this analysis, because we do not have yet a good *truth table* of galaxies observed in better conditions than DENIS in the NIR bands, and with simulated images we can obtain, in principle, as good statistics as we wish.

## 2. DETECTION METHODS

To increase the signal-to-noise ratio, we use a detection method applied to *smoothed* images. The algorithm is as follows: 1) *Smooth* the image with a given smoothing filter of size $N_{sm}$. 2) *Threshold* the smoothed image at a given $S/N$ above background. 3) Require a minimum number $N_{pix}$ of *connected pixels*. 4) Compare the pixel histogram around a detected object with pixel histograms with no objects, both with given aperture $ap$. This fourth step will use different statistical measures for the histogram comparison, each yielding reliability indices on the objects detected with steps 1–3.

Objects are detected in various steps: 1) Point sources are detected (at the other data analysis center in Leiden). 2) High $S/N$ galaxies (ellipticals, edge-on spirals) are detected with low smoothing scale with steps 1–3. 3) Low $S/N$ objects (face-on late-type spirals, planetary nebulae, etc.) are detected with large smoothing scale: a) with a large value of $N_{pix}$, thus limited to relatively bright, but highly reliable objects (steps 1–3); b) with a lower value of $N_{pix}$, giving less reliable, but more numerous candidates, on which we apply step 4) to weed out the false detections caused by noise.

## 3. PARAMETER SPACE

There are 4 different types of parameters: 1) *Object parameters*: object magnitude, type, and inclination. 2) *Image parameters*: seeing (atmospheric+instrumental), background (atmospheric+instrumental), read-out noise, instrumental gain. 3) *Field parameters*: stellar confusion, and interstellar extinction. 4) *Algorithmic Parameters*: smoothing length $N_{sm}$, threshold $S/N$, minimum number of connected pixels $N_{pix}^{min}$, and histogram aperture $ap$.

Once we fix the algorithm parameters, we use the image parameters, given by the processing of the raw images, and estimate the field parameters (from the galactic coordinates and from the total number of point sources detected). Of course, we cannot know in advance the object parameters for the detection, but use them for the selection functions.

## 4. RESULTS AND CONCLUSIONS

We begin by applying the first 3 detection steps to simulated images with and without objects (we focus here on the most difficult detection case, that of late-type face-on spirals, which we model as exponential disks). Figure 1 shows portions of simulated images without (*left*) and with (*right*) boxcar smoothing ($N_{sm} = 11$). The plots illustrate the low signal-to-noise ratio of objects and the importance of smoothing the image. The fainter galaxies are detected but are indistinguishable from noise.

We searched for the optimal 95% completeness magnitude limits over the algorithmic parameter space, using 20 objects per magnitude bin. Table 1 shows the results for the standard image parameters ($\mu_{K'}^{\rm sky} = 13.1$, PSF $= 1''$ [FWHM] and $\sigma_{\rm bg} = 4\,{\rm ADU}$), allowing for 0 and 20 (*asterisks*) false detections, using the optimal boxcar and gaussian filters. $K'_{\rm lim}$ depends sensitively on the background level, increasing with $\mu_{K'}^{\rm sky}$ (instead of as $0.5\mu_{K'}^{\rm sky}$ as for point sources). $K'_{\rm lim}$ depends little on the PSF (because of the large pixel size: $3''$), and also little on the instrumental gain (or on $\sigma_{\rm bg}$ in ADU). It is of course necessary to determine the background and its dispersion to high accuracy.

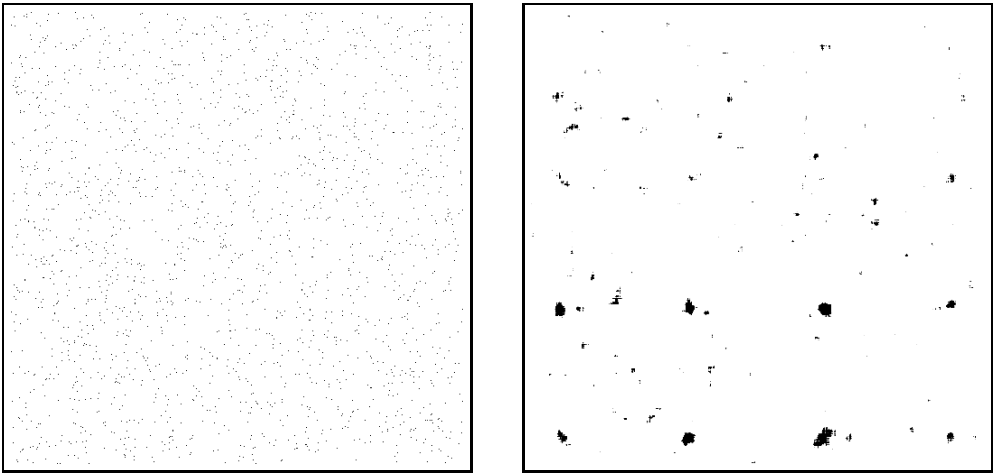

Figure 1: *Left:* Portion of raw DENIS-like simulated image. *Right:* Same as left, but smoothed with an $N_{\rm sm} = 11$ boxcar filter. Both figures are thresholded at $2.5\,\sigma$. Sixteen face-on disk galaxies are placed on a $4 \times 4$ square grid. Objects in a given row have the same total magnitude, and the objects in higher rows are increasingly fainter.

Table 1: Completeness limits

| Filter | $N_{\rm sm}$ | $\nu$ | $N_{\rm pix}^{\rm min}$ | $K'_{\rm lim95}$ | $N_{\rm pix}^{\rm min*}$ | $K'^{*}_{\rm lim95}$ |
|---|---|---|---|---|---|---|
| Boxcar | 11.0 | 2.5 | 63 | 12.4 | 36 | 13.1 |
| Gaussian | 1.5 | 3.0 | 42 | 12.4 | 4 | 13.1 |

We have tested this method with different filters (boxcar, gaussian, median, filter with pixel rejection or replacement ...). The results in Table 1 show that the boxcar filter gives similar completeness limits as the gaussian filter (other filters are not better): because of the very low $S/N$ of the objects, the selection function is, to first approximation, independent of the shape of the filter! Hence simple filters such as the boxcar or the gaussian are preferred.

**Acknowledgments** — We thank I. Vauglin for her oral presentation of this work.